\documentclass[a4paper]{article}

\usepackage{icrc2013}
\usepackage{amsmath}
\usepackage{float}

\title{Argentinian multi-wavelength scanning Raman lidar to observe night sky atmospheric transmission}

\shorttitle{Argentinean multi-wavelength scanning Raman lidar}

\authors{
Juan Pallotta$^{1}$,
Pablo Ristori$^{1}$,
Lidia Otero$^{1}$,
Fernando Chouza$^{1}$,
D'El\'{i}a Ra\'{u}l$^{1}$,
Francisco Gonzalez$^{1}$,
Alberto Etchegoyen$^{2}$,
Eduardo Quel$^{1}$
for the CTA consortium.
}

\afiliations{
$^1$ UNIDEF-CEILAP-(CITEDEF-CONICET), UMI-IFAECI-CNRS (3351) - Buenos Aires, Argentina. \\
$^2$  ITeDA (CNEA – CONICET - UNSAM) - Buenos Aires, Argentina. \\
}

\email{pablo.ristori@gmail.com}

\abstract{This paper discusses the multi-wavelength scanning Raman lidar being built at Lidar Division, CEILAP (CITEDEF-CONICET) in the frame of the Argentinean Cherenkov Telescope Array (CTA) collaboration to measure the spectral characteristics of the atmospheric aerosol extinction profiles to provide better transmission calculations at the future  CTA site. This lidar emits short laser pulses of 7-9 ns at 355, 532 and 1064 nm at 50 Hz with nominal energy of 125 mJ at 1064 nm. This wavelengths are also used to retrieve the atmospheric (air, aerosol and clouds) backscattered radiation in the UV, VIS and IR ranges. Raman capabilities were added in the UV and VIS wavelengths to retrieve the spectral characteristics of the aerosol extinction and the water vapor profile. Due to the expected low aerosol optical depth of the future site, the short observation period as well as the extension of the observation, an enhanced collection area is required. This system uses six 40 cm f/2.5 newtonian telescopes to avoid dealing with bigger mirror deformation, aberration issues and higher costs that imply the use of a single mirror with the same collection surface. In addition, dismounting of single mirrors for replacement or recoating will affect slightly the performance but not the operation. The additional alignment procedure has been solved by an automatic mirror alignment to follow the line of sight of the observation during the acquisition period. The system was designed to operate in hard environmental conditions, as it is completely exposed to the outside weather conditions, when its shelter is  fully opened to provide  360$^\circ$ observations.}

\keywords{lidar, aerosols, atmosphere, CTA.}

\begin{document}
\maketitle

\section{Introduction}

The range-dependent spectral atmospheric transmission in the line of sight of the telescopes is of major interest to CTA. In this sense, multi-wavelength scanning Raman lidars are able to acquire this information fast and accurately \cite{bib:matthias}. While these systems are routinely used in regional networks to provide information about aerosol extinction and vertical distribution \cite{bib:matthias, bib:murayama, bib:bosenberg}, CTA observation conditions are  different and need special considerations. The most challenging  requirements  are related to the scanning  capabilities and the fast acquisition time. Most of the systems provide vertical observation profiles, which can be done placing the lidar in a clean and thermally stabilized room with a rooftop aperture simplifying the system construction. Scanning lidars  are out in the open exposed to the outside weather conditions, but also needs this kind of protection to its critical parts (laser and laser optics) when they are measuring and the whole instruments needs a shelter, when not in use. Acquisition time is basically a function of laser energy, repetition rate, collection surface and reception optical efficiency, which must be improved to attain accurate profiles in a short period of time. In addition, these lidars are intended to be used remotely by an operator without an a priori knowledge of lidar techniques. Therefore the operation has to be simple, with most of the measurement specific processes running in a hidden layer.

\section{Requirements for detection wavelength range of the Lidar system}
Due to the fact that aerosol layers increase the backscattered radiation at the lidar emission wavelength, most of these lidars (called ‘elastic’ lidars) are able to detect the presence of even small aerosol layers. For these systems, the lidar return can be expressed as follows:

\begin{equation}
P(r, \lambda_E)=  K G(r) r^{-2} \beta(r, \lambda_E, \lambda_D) T(r, \lambda_E ) T(r, \lambda_D)
\label{eq_lidar}
\end{equation}

in which \textit{K} is a constant that takes into account terms such as the laser energy, the collection surface and the overall system efficiency; \textit{G(r)} is the fraction of the light collected by the telescope that is sent to the detector (overlap function),  $\beta(r, \lambda_E, \lambda_D)$ is the atmospheric backscatter of the laser wavelength $\lambda_E$ at the detected wavelength $\lambda_D$,  \textit{T(r, $\lambda_E$)} and  T(r, $\lambda_D$ ) are the upwards laser transmission up to the the range \textit{r} and downwards  backscattered transmission to the lidar’s telescope defined as:

\begin{equation}
T(r, \lambda ) =  exp(-\int_0^{r}{ (\alpha^{m}(r, \lambda) + \alpha^{p}(r, \lambda) + \alpha^{abs}(r, \lambda) }) dr)
\label{eq_transmission}
\end{equation}

where $\alpha^{m}(r, \lambda)$ is the extinction coefficient for molecules, $\alpha^{p}(r, \lambda)$ is for particles, and $\alpha^{abs}(r, \lambda)$ for absorbing species. An extensive discussion of the lidar equation can be found in \cite{bib:pappalardo}. In the presence of a non-negligible aerosol layer it is more convenient to measure the backscattered return at a wavelength free of aerosol backscatter. This can be done by measuring the backscattered vibro-rotational nitrogen or oxygen Raman lines, pure rotational Raman spectra or by measuring using high spectral resolution lidars. The first of these three options is the simplest to achieve. The inversion equation is:

\begin{align}
\alpha^{p}(r, \lambda_E)=  \frac{d}{dr}  \ln\left\{ \frac{N(r)}{r^{2}P(r,\lambda_v)} \right\} - \alpha^{m}(r, \lambda_E) \nonumber \\ -\alpha^{m}(r, \lambda_{\nu})  - \alpha^{p}(r, \lambda_{\nu})
\label{eq_inv_raman}
\end{align}

where \textit{N} is the number of molecules and the sub-index $\nu$ stands for the Raman shift produced by the selected molecule (backscattered returns are red-shifted by 2331 $cm^{-1}$ and 1556 $cm^{-1}$ by the nitrogen and oxygen molecules respectively). The fourth term which is unknown can be estimated using the \r{A}ngstr\"{o}m relation:

\begin{equation}
\alpha^{p} (r, \lambda_E)= \alpha^{p} (r, \lambda_{\nu})  \left\{ \frac{\lambda_E}{\lambda_{\nu}} \right\}^{\r{A}(r)}
\label{eq_angstrom}
\end{equation}

in which \r{A}(r) stands for the \r{A}ngstr\"{o}m range-dependent coefficient that can be modeled or derived from a second elastic Raman wavelength  \cite{bib:otero, bib:mattis}. As Cherenkov telescopes detect polychromatic light, aerosol atmospheric transmission for a single wavelength may not be appropriate to characterize effects of a given air mass so the wavelength dependency of the aerosol transmission must be considered.
From these requirements the Argentinean CTA collaboration decided to build a multiwavelength Raman lidar emitting in the fundamental, second and third harmonics of a Nd:YAG laser (1064nm, 532nm and 355nm) and receiving these wavelengths and the nitrogen Raman shifted wavelengths from 532 nm (607 nm) and 355 nm (387 nm). A sixth wavelength at 408 nm was also used to detect the water vapor Raman return. The way that the knowledge of these parameters impacts over Cherenkov telescopes is described in \cite{bib:doro, bib:garrido}.

\section{Lidar design}

\begin{figure}[h!]
  \centering
  \includegraphics[width=0.4\textwidth]{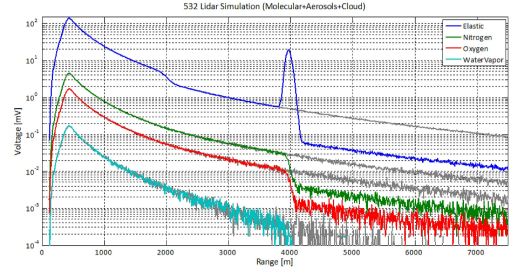}
  \caption{Atmospheric lidar return from an emission source of 100 mJ, 10 Hz repetition rate at 532 nm acquired by a 45 cm f/3 telescope during 15 minutes.}
  \label{simulacion_fig}
 \end{figure}

Lidar simulations can provide a first estimate of the lidar profiles for a known atmosphere. As an example, \ref{simulacion_fig}  shows the simulated elastic signal (532 nm) and Raman returns for a well-mixed boundary layer in presence of mid altitude clouds, measured with a 40 cm diameter, 100 cm focal length Newtonian telescope, emitting with a Nd:Yag laser (100 mJ @ 532 nm).

The polychromator efficiency and the PMT quantum efficiency are 30\% and 25\% respectively. Elastic signal (in blue, attenuated 100 times) and Raman signals (in green, red and cyan for nitrogen, oxygen and 100\%  humidity water vapor, respectively) correspond to a US Standard Atmosphere profile with ground pressure of 1013.25 Pa and temperature of 15$^\circ$C and an adiabatic lapse rate of -6.5 K/km. The aerosols in the atmospheric boundary layer are well mixed up to a height of 2 km and an entrainment region of 100 m. The simulation is being performed with and without the presence of a cloud at 4km.

To increase the lidar signal it is more effective to increase the telescope collection surface. This can be done by increasing the telescope diameter of a single mirror or increasing the number of mirrors. The first option was chosen by the Institut de F\'{i}sica d’Altes Energies (IFAE) and the Universitat Autonoma de Barcelona (UAB), located in Barcelona (Spain) and the Laboratoire Univers et Particules de Montpellier (LUPM) in Montpellier (France); while the second option was chosen by Centro de Investigaciones en Láseres y Aplicaciones (CEILAP) in Villa Martelli, Buenos Aires (Argentina). Our reason to select the second option was the possibility to use standard 40 cm diameter, f/2.5 parabolic mirrors and 1 mm optical fibers (NA=0.22) to transfer the collected light to the polychromator. This solution permits to extract any mirror for being recoated or exchanged keeping the other five mirrors in the system with a total system efficiency of 83 \%. Furthermore mirror construction and coating can be done by standard methods. Conversely an equivalent 98 cm diameter, f = 1 m single parabolic mirror telescope is difficult to create.
The main drawback of the chosen solution is that each of the six telescopes must be aligned properly to attain maximum system efficiency. That was the main subject that focused our attention.

\section{Prototype lidar construction}

\begin{figure}[h!]
  \centering
  \includegraphics[width=0.4\textwidth]{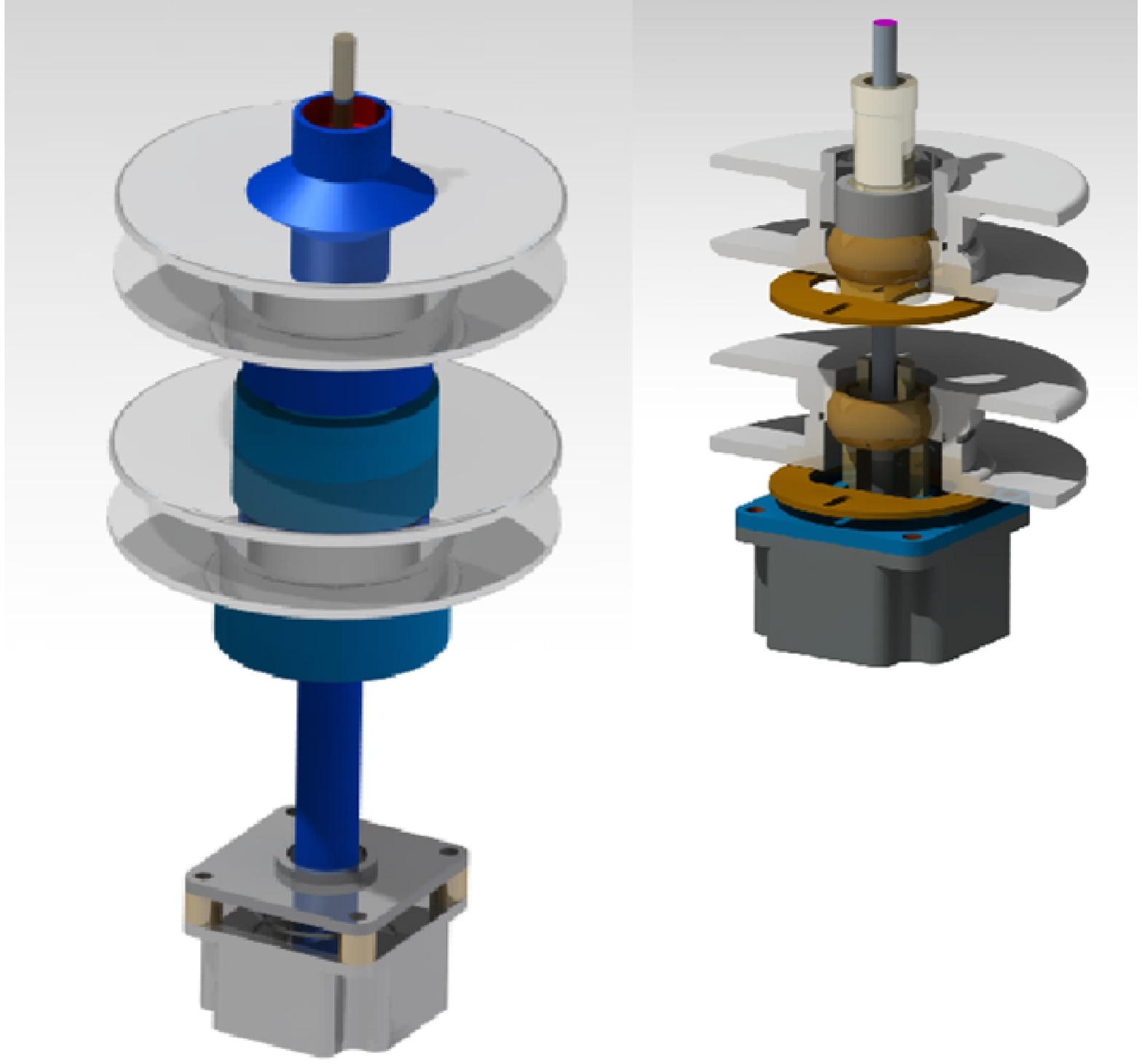}
  \caption{Left: first prototype with differential screws and mechanical parts on demand to provide smoother movements. Right: final version with normal screws and more commercial parts to achieve the same movements as the first prototype.}
  \label{actuadores_fig}
 \end{figure}

In the deployment process of this lidar we have designed first prototypes for every mechanical and electronic part. Each prototype was tested and redesigned to be improved if necessary.
This was the case of the two actuators needed to align a single mirror. The original actuators were based on differential screws and were different depending on the required degrees of freedom. The first tests were done using these actuators. With this experience, the final version was simpler, lighter, more stable and using commercial ball joints at every articulated part. A single configuration was used for both actuators, while their degrees of freedom were reduced/increased by tightening/untightening internal screws. Reducing the number of steps it was possible to replace the differential screws to standard ones. Figure \ref{actuadores_fig} shows a scheme of both actuators at the same scale.

\begin{figure}[h!]
  \centering
  \includegraphics[width=0.4\textwidth]{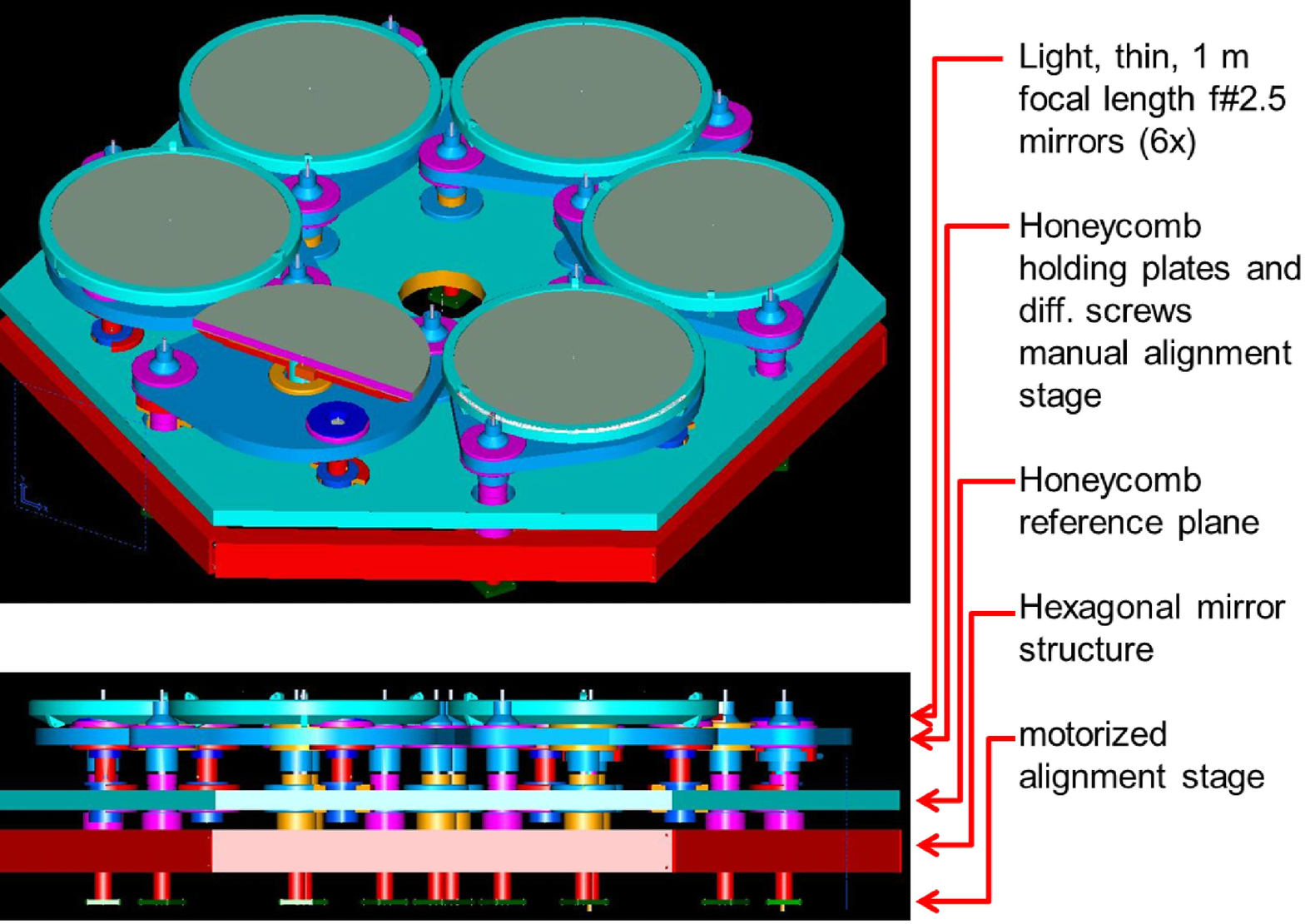}
  \caption{Multi-mirror setup of the lidar system from the top and lateral view showing the different stages of the system.}
  \label{telescopios_fig}
 \end{figure}

The multi-mirror telescope unit was designed to provide a maximum stability to the system with a minimum weight. While honeycomb was used for the multi-mirror reference plane, carbon fiber tubes were used to place the optical fiber at the mirror focal plane. Nylon pieces were synterized at the end of the carbon tubes to provide better fixation. The resulting setup (without carbon tubes) is shown on Figure~\ref{telescopios_fig}.

\begin{figure}[h!]
  \centering
  \includegraphics[width=0.4\textwidth]{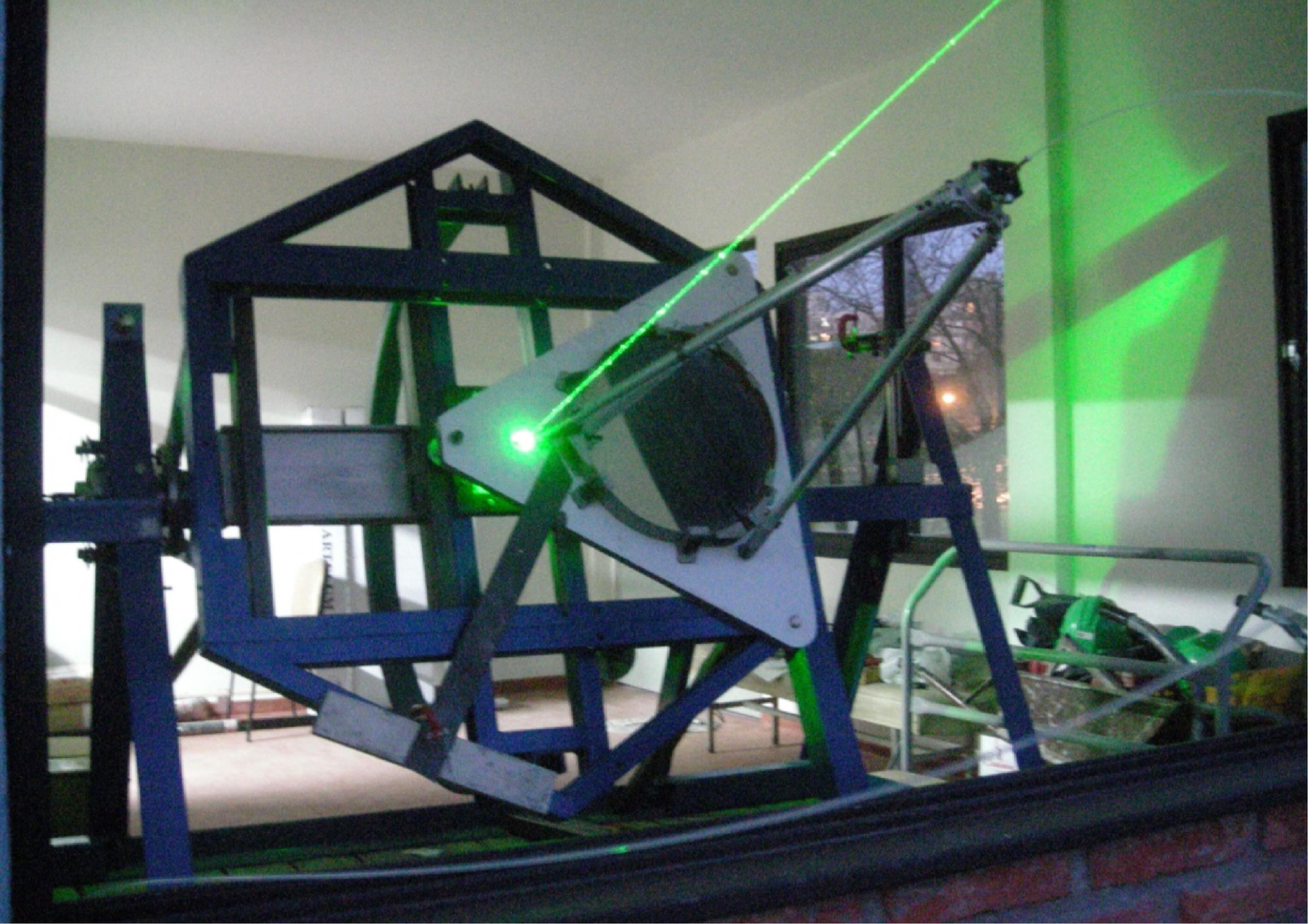}
  \caption{First test of the lidar mechanical structure. A single mirror was used combined with the laser source. The mirror structure was made for this experiment.}
  \label{primeras_seniales_fig}
 \end{figure}

\begin{figure}[h!]
  \centering
  \includegraphics[width=0.4\textwidth]{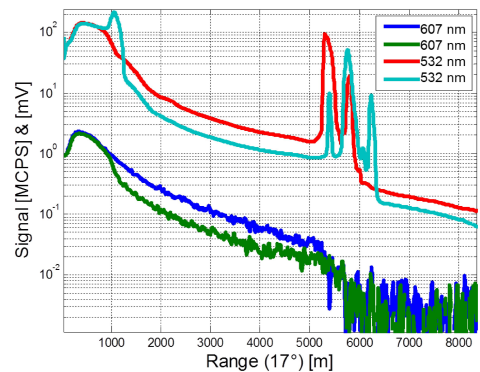}
  \caption{Two different datasets for 532 nm and the correspondent Nitrogen Raman wavelength return signal. It can be noticed the presence of several cloud layers in the elastic channel and the efficiency of the Raman channel to block the 532 nm returns in 607 nm channel.}
  \label{seniales_fig}
 \end{figure}

In a first stage the lidar structure was studied with a single mirror and a laser emission to test its mechanical behavior during lidar operation. Figure \ref{primeras_seniales_fig} shows the lidar as it was operated and Figure \ref{seniales_fig} the retrieved lidar profiles.

\section{New shelter-dome}

\begin{figure}[h!]
  \centering
  \includegraphics[width=0.4\textwidth]{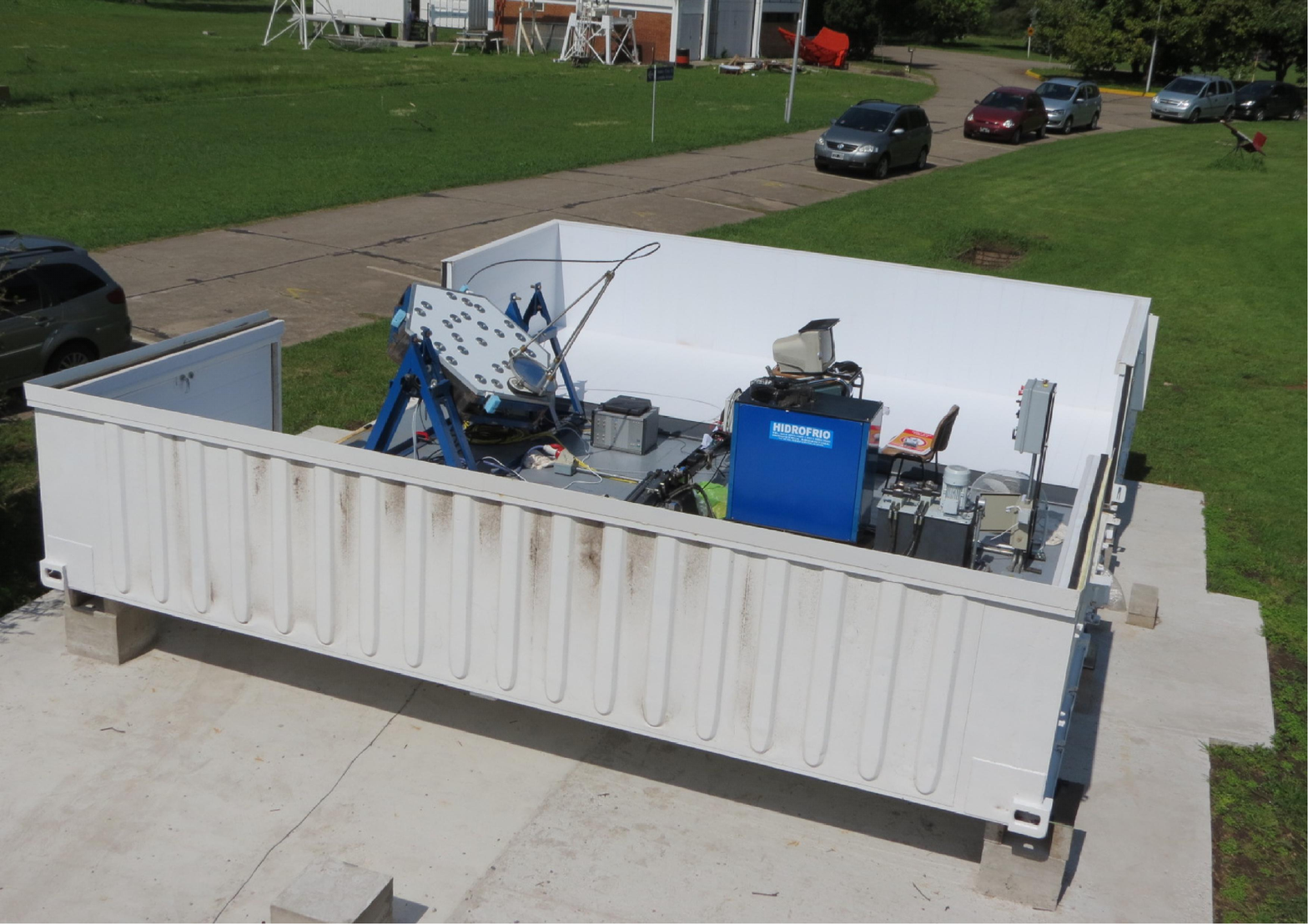}
  \caption{Lidar instrument inside the shelter based on a standard 20 feet container reinforced and adapted to be opened by two hydraulic pistons at CEILAP.}
  \label{open_shelter}
 \end{figure}

During 2012 a new shelter-dome to host the multiangle lidar was acquired. Based on the idea of CLUE shelter, we have adopted the same concept to host the lidar. Based on the experience of our colleagues in Barcelona and Montpellier we have reinforced a standard 20 ft. shelter and we have equipped it with hydraulic pistons conceived to open and close it automatically as is shown on Figure \ref{open_shelter}. 
These and other operations of the container will be included in the lidar control software to automatize it for remote operation this year.

\section{Software}

\begin{figure}[h!]
  \centering
  \includegraphics[width=0.4\textwidth]{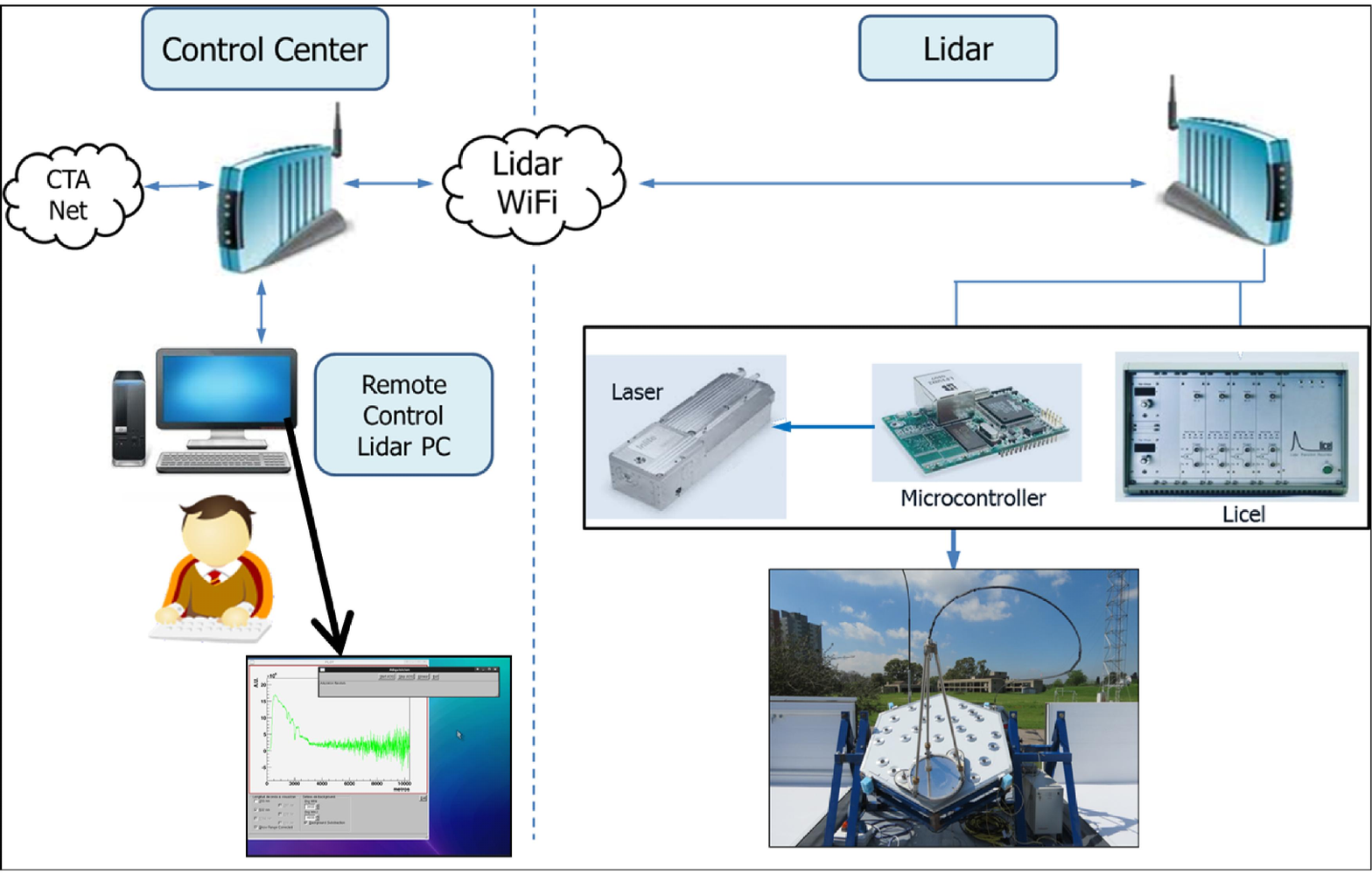}
  \caption{Schematic diagram of the remote control proposed for multiangle lidar system.}
  \label{open_shelter_fig}
 \end{figure}

The whole multi-angle lidar is actually remotely controlled using a Wi-Fi link from the control PC to the lidar shelter creating a local lidar network under the TCP/IP protocol. Both acquisition and shelter controls will be operated remotely by the shifter. This control system is fully functional. The operation scheme is presented in Figure \ref{open_shelter_fig}.

This kind of link has several advantages due to its data bandwidth, simplicity and modularity. To accomplish that, WiFi routers with Wireless Distribution System (WDS) feature are used, with the lidar-side router working as access point (AP). This AP distributes the lidar network data to the electronics: a Licel transient recorder rack (LICEL transient recorder TR20-160, with Ethernet interface) and a microcontroller. A MAC filter is set for both routers to ensure a reliable way to limit the endpoints connected to the lidar network.

Acquisition process is developed in C/C++ and ROOT libraries, and the control of the shelter (open/close, turn on/off devices, etc.) was built with a HTML interface.

The automatic alignment for multiple telescopes is performed also using this program. From the lidar side a built-in Ethernet interface and integrated TCP/IP stack microcontroller acts as interface to control electronic devices.

\section{Work in progress and future plans}

We currently develop a new concept design for the scanning system in collaboration of the Mechanical Department of CITEDEF. This structure that holds the hexagonal honeycomb and structure below (laser, motors and drivers) is much more compact than the existing one. We have planned the following tasks in the near future:

\begin{itemize}
\item Use the complete reception system with the six telescopes in simultaneous alignment and measuring operation. 
\item Build spectrometric 6 lines (3 elastic + 3 Raman) spectrometric box using the same optical configuration of the four other lidars being built at CEILAP and a fifth lidar operating since October 2012 at Comodoro Rivadavia. The most important difference will be its reduced size and optical fiber coupling.
\item Build the motorized azimuthal-zenithal mechanism, providing the system with scanning capabilities. 
\item Place the lidar with the motorized structure on the shelter to reach full operation condition of the system.
\end{itemize}

\section{Conclusions}

The construction of the presented multiwavelength scanning Raman lidar will be able to provide spectrally-resolved aerosol extinction profiles to characterize the atmospheric transmission at any required line of sight and in a short period of time. The modularity of the telescope system will permit the system maintenance and optimization while being operated reducing non-operational times. The collaboration of CEILAP, IFAE/UAB and LUPM to improve their lidar systems will permit to attain the requested goals in terms of system construction, lidar testing, instrumentation control and lidar signal processing.

\vspace*{0.5cm}

\footnotesize{{\bf Acknowledgment:}{Authors wish to thank CITEDEF main workshop’s technicians, Mario Proyetti and José Luis Luque from the CEILAP workshop for their support on this development. We gratefully acknowledge support from the following agencies and organizations: Ministerio de Ciencia, Tecnolog\'ia e Innovaci\'on Productiva (MinCyT), Comisi\'on Nacional de Energ\'ia At\'omica (CNEA) and Consejo Nacional  de Investigaciones Cient\'ificas y T\'ecnicas (CONICET) Argentina; State Committee of Science of Armenia; Ministry for Research, CNRS-INSU and CNRS-IN2P3, Irfu-CEA, ANR, France; Max Planck Society, BMBF, DESY, Helmholtz Association, Germany; MIUR, Italy; Netherlands Research School for Astronomy (NOVA), Netherlands Organization for Scientific Research (NWO); Ministry of Science and Higher Education and the National Centre for Research and Development, Poland; MICINN support through the National R+D+I, CDTI funding plans and the CPAN and MultiDark Consolider-Ingenio 2010 programme, Spain; Swedish Research Council, Royal Swedish Academy of Sciences financed, Sweden; Swiss National Science
Foundation (SNSF), Switzerland; Leverhulme Trust, Royal Society, Science and Technologies Facilities Council, Durham University, UK; National Science Foundation, Department of Energy, Argonne National Laboratory, University of California, University of Chicago, Iowa State University, Institute for Nuclear and Particle Astrophysics (INPAC-MRPI program), Washington University McDonnell 
Center for the Space Sciences, USA. The research leading to these results has received funding from the European Union's Seventh Framework Programme ([FP7/2007-2013] [FP7/2007-2011]) under grant agreement nÂ° 262053.}}

\end{document}